\documentclass[aps,prl,showpacs,twocolumn,superscriptaddress,amsmath,amssymb,floatfix]{revtex4-1}
\usepackage{tabularx,wasysym}
\usepackage{bm}
\usepackage{epsfig,subfigure}
\usepackage{multirow}
\usepackage{graphicx}
\usepackage{color}
\usepackage{amsfonts}
\usepackage{exscale}
\usepackage{amsbsy}
\graphicspath{{Fig/}}
\usepackage[normalem]{ulem} 
\usepackage{soul}
\usepackage[colorlinks,urlcolor=blue,linkcolor=blue,anchorcolor=blue,citecolor=blue]{hyperref}

\begin{document}
\title{Kagome chiral spin liquid as a gauged $U(1)$ symmetry protected topological phase}
\author{Yin-Chen He}
\affiliation{Max-Planck-Institut f\"{u}r Physik komplexer Systeme, N\"{o}thnitzer Str. 38, 01187 Dresden, Germany}
\author{ Subhro Bhattacharjee} 
\affiliation{Max-Planck-Institut f\"{u}r Physik komplexer Systeme, N\"{o}thnitzer Str. 38, 01187 Dresden, Germany}
\affiliation{International Centre for Theoretical Sciences, Tata Institute of Fundamental Research, Bangalore 560012, India}
\author{Frank Pollmann}
\affiliation{Max-Planck-Institut f\"{u}r Physik komplexer Systeme, N\"{o}thnitzer Str. 38, 01187 Dresden, Germany}
\author{R. Moessner}
\affiliation{Max-Planck-Institut f\"{u}r Physik komplexer Systeme, N\"{o}thnitzer Str. 38, 01187 Dresden, Germany}
\date{\today}

\begin{abstract}
While the existence of a chiral spin liquid (CSL) on a class of spin-1/2 kagome antiferromagnets is by now well-established numerically, a controlled theoretical path from the lattice model leading to  a low energy topological field theory is still lacking. This we provide via an explicit construction, starting from reformulating a microscopic model for a CSL as a lattice gauge theory, and deriving the low-energy form of its continuum limit. A crucial ingredient is the realisation that the bosonic spinons of the gauge theory exhibit a $U(1)$ symmetry-protected topological (SPT)  phase, which upon promoting its $U(1)$ global symmetry to a local gauge structure (``gauging") yields the CSL. We suggest that such an explicit lattice-based construction involving gauging of an SPT phase can be applied more generally to understand topological spin liquids. 
\end{abstract}
\maketitle
\paragraph{Introduction:} Quantum spin liquids (QSLs)  are long-range quantum entangled phases of interacting spins that support fractionalized excitations--generally referred to as spinons \cite{Anderson1973,Wen1990,Wen_book,Fradkinbook}.  The task of predicting and controllably describing a QSL state relevant to a particular lattice spin model remains a challenge to date. This is all the more important at present in the context of understanding lattice Hamiltonians relevant for material \cite{Han2012, Clark2013}. Based on our current theoretical understanding of QSLs as deconfined phases of effective gauge theories that emerge as low energy descriptions of spin models, answering the above questions require addressing two key issues: (i) identifying a faithful low energy lattice gauge theory (LGT), 
 for a given microscopic spin model; (ii) showing that this lattice gauge model exhibits a QSL as a deconfined phase supporting fractionalized excitations.

An example of such a controlled LGT description of QSL  is the quantum dimer model on the triangular lattice \cite{Moessner2001}, which can be mapped to a  $ Z_2$ gauge theory \cite{Fradkin1979,Moessner2001b}, thereby potentially realizing a $ Z_2$ QSL \cite{Moessner2001, Balents2002,Kitaev2006}.
However, systematic implementations of such a construction to obtain other QSLs in two dimensional magnets, such as chiral \cite{Kalmeyer1987,Wen1989,Yang1993} or critical  QSLs \cite{Wen_book,Fradkinbook} have proven difficult.
These QSLs can be obtained using LGT with fermionic spinons \cite{Wen_book, Fradkinbook}. However, how to obtain such fermionic spinons in a controlled fashion is largely an  open question.
On the other hand a faithful LGT with bosonic spinons and a compact $U(1)$ gauge field can be obtained for some microscopic models,  {\it e.g.}, quantum dimer model on bipartite lattices \cite{Rokhsar1988,Fradkin1990,Fradkinbook} or spin models on the checker-board lattice \cite{Shannon2004}.
 At low energies these typically result in  pure compact $U(1)$ gauge theories \cite{Fradkinbook}, for which, in two spatial dimensions, the LGT will be generically in the confining phase \cite{Polyakov} leading to conventional ordering (e.g., valence bond solid);
 but in three spatial dimension, it can host a stable QSL phase \cite{Moessner2003,Hermele2004,Castelnovo2008}.

An interesting situation occurs in the low energy limit of easy-axis spin-1/2 kagome antiferromagnets. In this case, the low energy physics  is described by a compact $U(1)$ LGT   \emph{coupled to dynamical bosonic spinons}  \cite{Nikolic2005,Nikolic2005b} (in contrast to the above mentioned dimer models, which are described by \emph{pure} compact LGT).
The dynamical bosonic spinons carry finite gauge charge and their presence can  have drastic influences on the system \cite{Nikolic2005,Moessner2000}. 
In addition, recent numerical simulations \cite{He2015} show that such systems can stabilize QSLs, including the enigmatic kagome spin liquid \cite{Yan2011, Ran2007,Depenbrock2012,Jiang2012,Iqbal2013} as well as a chiral spin liquid (CSL) \cite{Kalmeyer1987,Wen1989} over large parameter regimes. The latter is a gapped QSL that breaks time reversal symmetry (spontaneously or explicitly), exhibits topological ground state degeneracy when put on a torus and supports gapless chiral edge states with quantized (fractional) spin-Hall conductivity \cite{Kalmeyer1987,Wen1989}.
How can such kagome CSL be understood and described from the point of view of the above $U(1)$ LGT? 
(One possible way to understand the CSL is through a lattice Chern-Simons theory on the kagome lattice \cite{Kumar2014, Kumar2015}.)
This is the fundamental question that we will formulate an answer to.

In this Letter, taking clues from recent developments in symmetry protected topological phases (SPTs) \cite{Haldane1983,Pollmann2010,Chen2010,Chen2013}, we  explicitly construct a LGT description of the CSL phase and obtain its continuum limit a controlled way. 
Unlike QSLs, SPTs have no intrinsic topological order (and therefore no fractionalized bulk excitations), but support symmetry protected anomalous gapless \cite{Levin2012} or topologically ordered edges \cite{Vishwanath2013}. 
We explore the idea of  ``gauging" an SPT to obtain a topologically ordered phase \cite{Levin2012,Barkeshli2013}. 
This means promoting the global symmetry that protects a given SPT to a local gauge structure, will yield a topologically ordered phase.
In particular, we derive a controlled description of the CSL  as a gauged $U(1)$ SPT (bosonic integer quantum Hall state) \cite{Senthil2013, Lu2012,He2015b}. 
We implement the idea for two microscopic easy-axis kagome spin models, which (or similar versions) were recently shown to host a CSL by means of numerical simulations (DMRG \cite{DMRG}) \cite{He2014,ssgong14,Bauer2014,He2015}.

\begin{figure}
\includegraphics[width=0.39\textwidth]{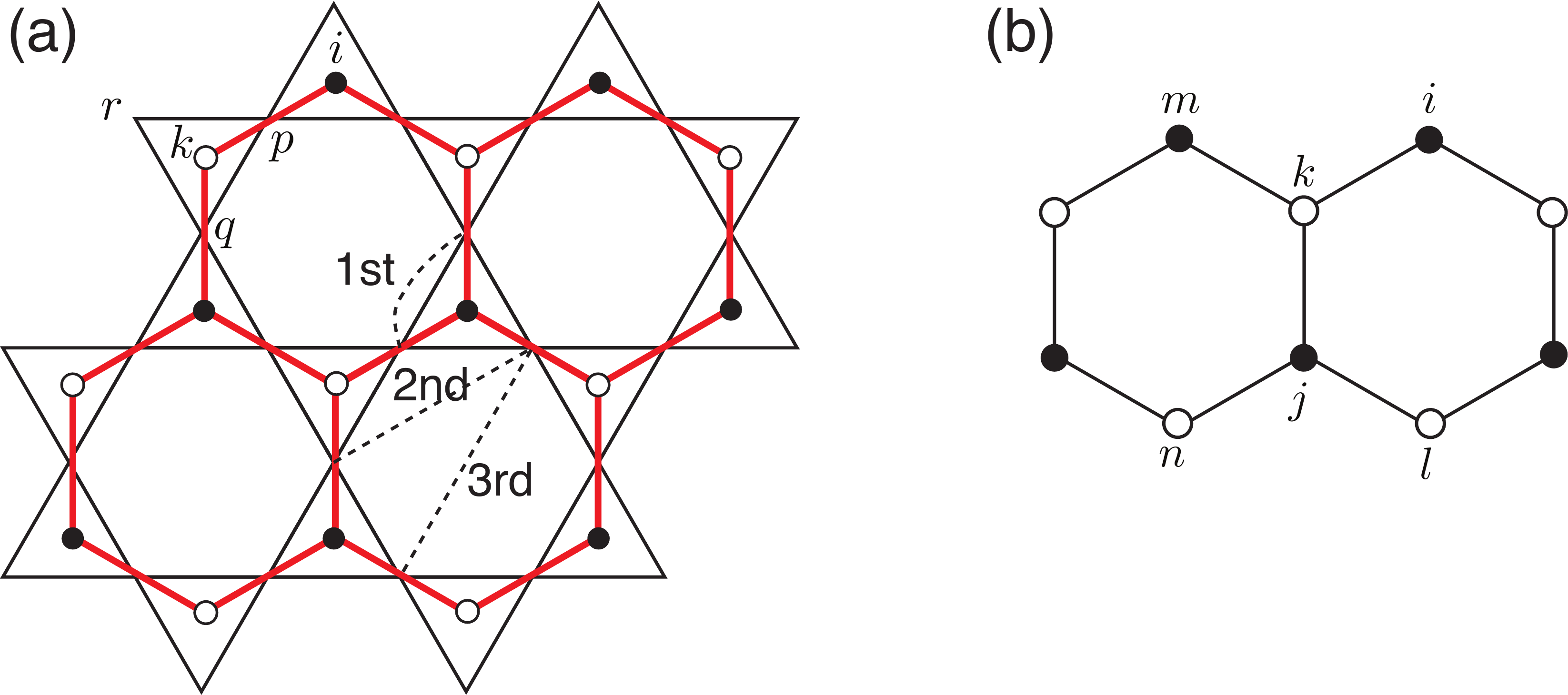}\caption{\label{fig:kagome_honeycomb} (a) Kagome lattice and medial honeycomb lattice. (b) The lattice gauge theory is defined on a honeycomb lattice.
The gauge field is only defined on the first neighbor link, hence the gauge field on other links, {\it i.e.}, second neighbor is $\mathcal A_{ij}=\mathcal A_{ik}+\mathcal A_{kj}$.}
\end{figure}

\paragraph{Lattice gauge theory:}
We illustrate our main ideas for the chiral spin-1/2 easy-axis model on the kagome lattice described by the Hamiltonian 
\begin{align}
H_{\text{chiral}}= J_z \sum_{\langle p q\rangle} S_p^z S_q^z+\lambda\sum_{p,q,r\in \bigtriangledown, \bigtriangleup} \vec S_r \cdot (\vec S_p \times \vec S_q),  \label{eq:Ham_B}
\end{align}
with $J_z\gg \lambda>0$, which is different from a previous work \cite{Bauer2014} that studied case  $J_z=0$. The three-spin term  is the \emph{scalar spin chirality}, which breaks time-reversal and parity explicitly, and has been used to engineer a CSL before \cite{Schroeter2007,Bauer2014}.  After discussing the general structure of our construction, we apply it to  the CSL phase of the generalized $XXZ$ model on the kagome lattice \cite{He2015} towards the end of the paper.

In the classical Ising limit ($\lambda=0$) of Eq.~(\ref{eq:Ham_B}), the ground state manifold has an extensive degeneracy and is given by all classical spin configurations that fulfill %
\begin{equation}
\sum_{p \in \bigtriangleup,\bigtriangledown} S_p^z=\pm 1/2 \label{eq:cons}
\end{equation}
for each triangle of the kagome lattice. 
In this manifold, the three-spin terms act as a perturbation which lifts the classical degeneracy by forming a coherent superpositions of classical configurations. 
We next formulate the resulting degenerate perturbation theory \cite{Nikolic2005}. This is conveniently done 
 in terms of degrees of freedom  that live at the center of each triangle; these centers form a honeycomb lattice:
\begin{align}
\sum_{ p \in \bigtriangleup_i} S_{ p}^z  =a^\dag_i a_i-\frac 1 2, ~~~  \sum_{ p \in \bigtriangledown_k} S_p^z =b^\dag_k b_k-\frac 1 2. 
\end{align}
Here, $a^\dag_i$, $b^\dag_k$ denote the creation operator for the hard-core boson on the $A$, $B$ sublattice of the honeycomb lattice (Fig. \ref{fig:kagome_honeycomb}).

We can now define a lattice electric field $ E_{ik}$ on the links of the honeycomb lattice
\begin{equation}
 E_{ik}=- E_{ki}=(S^z_{ p}+1/2),
\end{equation}
where $i(k) \in\bigtriangleup(\bigtriangledown)$, such that  Gauss' law is fulfilled on each site of the honeycomb lattice:
\begin{align}
 \sum_{k; i\in \bigtriangleup}  E_{i k}=n^a_{i}+1,~~~
\sum_{i; k\in \bigtriangledown}  E_{k i}=-(n^b_{k}+1).
\end{align}
The summation runs over the three neighbors on the honeycomb lattice.  Here $n_i^a=a^\dag_i a_i$,  $n_k^b=b^\dag_k b_k$, and $\pm1$ represent static background charges on the two sublattices. 
The  spin flip operators are then given by
\begin{equation}
S^+_p = \exp (i \mathcal A_{ik})  a_i^\dag  b_k^\dagger \ \ \mbox{and} \quad S^-_p = \exp (i \mathcal A_{ki})   a_i  b_{k},\label{eq:spin_spinon}
\end{equation}
where $\mathcal A\in [0,2\pi)$ is the vector potential conjugate to $E_{ik}$ (here we have soften the hard-core constraint of $E_{ik}$ as usually done in similar systems \cite{Fradkinbook,Fradkin1990,Castelnovo2008,Hermele2004}). 
Conservation of $S^z$ implies a conserved $U(1)$ charge;  and the $a^\dag$ and $b^\dag$ carry fractional $+1/2$ charges, hence are spinons.
We note that the ground state has vanishing magnetization $\sum S^z=0$, thus the spinons are effectively  at half filling (per site).
In addition, the two flavors of spinons carry opposite  $U(1)$ gauge charges of $\mathcal{A}$, {\it i.e.}, under gauge transformation: $a_i\rightarrow e^{i\theta_i}a_i$, $b_k\rightarrow e^{-i\theta_k}b_k$.

Note that the above mapping is  similar to the one used in quantum dimer models \cite{Fradkinbook,Fradkin1990} and quantum/classical spin ice \cite{Castelnovo2008,Hermele2004} with the crucial difference \cite{Nikolic2005},
that the present model has dynamical bosonic spinons in addition to the compact $U(1)$ gauge field. %

Using the above mapping we can now represent the effective microscopic model as a lattice gauge theory in terms of the hard-core bosons (spinons)  coupled to an emergent $U(1)$ gauge field (unimportant factors omitted): 
\begin{align}\label{eq:LGT_HamB}
H_{\textrm{chiral}}^{\textrm{LGT}}&=
\lambda\sum_{\langle\langle ij \rangle\rangle} \left[e^{i \mathcal  A_{ij}+i\pi/2} (2 n_k^b-1) a^\dag_i a_j +h.c.\right] \nonumber \\ &+ \lambda\sum_{\langle \langle kl \rangle\rangle} \left[e^{i\mathcal A_{lk}+i\pi/2}(2 n_j^a-1) b^\dag_k b_l + h.c.\right],
\end{align}
where we have the correlated hopping: that bosons (spinons) are hopping within the second neighbors (e.g. $\langle \langle i j \rangle\rangle$), and are coupled to the boson density in the intermediate site (e.g. $k$) as shown in Fig. \ref{fig:kagome_honeycomb}(b). 
%
In principle, there is also a Maxwell term for the gauge fields arising from ring-exchange around hexagons. 
However, since this is third order in perturbation theory ($\sim$$\lambda^3/J_z^2\ll\lambda$), we  neglect it. 

\paragraph{Gauge mean-field theory and the U(1) SPT state:} We now adopt a gauge mean-field (gMF) treatment \cite{Savary2012} to solve the above lattice gauge Hamiltonian Eq.~(\ref{eq:LGT_HamB}) and find the SPT phase advertised above.
The strategy is as follow:
We start by treating the dynamical gauge fluctuation as a mean-field gauge flux $\mathcal A \sim \langle \mathcal A\rangle=\mathcal A^0$ and consequently the local  $U(1)$ gauge structure is replaced by a $U(1)$ global symmetry. 
Then we solve the corresponding gMF Hamiltonian and find that the ground state is a gapped $U(1)$ SPT.
Finally we restore the local gauge invariance and  the gap protects the ground state against the gauge fluctuation. This gauged $U(1)$ SPT, as we shall show, is nothing but a CSL \cite{Barkeshli2013}.
To  completely define the gMF Hamiltonian, we need to determine the mean-field gauge flux $\sum_{\hexagon} \mathcal A^{0} $ for each hexagon.
As we neglect the Maxwell term, the dynamics of the matter field alone  determines the gauge flux.
In other words, the flux pattern for the gMF Hamiltonian that has the lowest energy is chosen.
Assuming that the solution preserves translation symmetry (as suggested by DMRG results), we consider two possibilities, $ \sum_{\hexagon} \mathcal A^{0}=0$ or $ \sum_{\hexagon} \mathcal A^{0}=\pi$. 
Using the iDMRG method \cite{McCulloch2008} we find that the ground state with the background flux  $ \sum_{\hexagon} \mathcal A^{0}=\pi$ (for each hexagon) has much lower energy  [$(E_0-E_\pi)/E_\pi\sim 10\%$].

With the background flux specified, the gMF Hamiltonian of Eq.~(\ref{eq:LGT_HamB}) describes bosons with correlated hopping subject to a time-reversal symmetry breaking flux of $\pi/2$ (the sum of the gauge flux $\mathcal A^0$ and the original hopping flux $\pm\pi/2$).
This is exactly the same model as studied in a previous work by us \cite{He2015b}, where we found that the ground state is a gapped $U(1)$ SPT phase with Hall conductance $\sigma_{xy}^c= 2$.
Phenomenologically, the correlated hopping terms in Eq.~(\ref{eq:LGT_HamB}) favor a mutual flux attachment \cite{Senthil2013} and thus stabilize  a $U(1)$ SPT phase.
In passing, we note that the correlated hopping term actually results from the projection of the scalar chirality to the effective lattice gauge model on the honeycomb lattice (see supplemental materials).
%

\paragraph{Chiral spin liquid:} The above gMF theory provides  insights to the correct continuum limit to describe the CSL in the original kagome spin model. This involves taking the low energy continuum theory of the $U(1)$ SPT and restoring the local $U(1)$ gauge invariance.  

The gMF Hamiltonian has two global $U(1)$ symmetries corresponding to 
(i) the overall particle/charge ($n_a+n_b$) conservation arising from the $S^z$ conservation of the original spin model;  (ii) the pseudospin ($n_a-n_b$) symmetry from the freezing of the gauge fluctuations, which  should hence be promoted back to a local gauge structure.
The $U(1)$ SPT is described by a condensation of composite bosons $\tilde a$, $\tilde b$, which is due to the mutual flux attachment: attaching flux of $b$ to the density of $a$ and vice versa \cite{Senthil2013}.
Those composite bosons $\tilde a$, $\tilde b$ carry the same quantum numbers as bosons $a$ and $b$, specifically $\tilde a$ carries $+1/2$ $S^z$ charge, and $+1$ gauge charge;  $\tilde b$ carries $+1/2$ $S^z$ charge, and $-1$ gauge charge.
The low energy theory including the gauge fluctuations is then given by the Lagrangian \cite{Senthil2013}
\begin{align}
\mathcal{L}=\mathcal{L}_a+\mathcal{L}_b+\mathcal{L}_{\rm int}+\mathcal{L}_{CS}+\mathcal{L}_{\mathcal{A}},
\end{align}
where
\begin{align}
\mathcal{L}_a=&i\tilde a^*\left[\partial_0-i\left(\frac{1}{2}A^{\rm ext}_0+\mathcal{A}_0\right)+i\alpha_0\right]\tilde a\nonumber\\
&-\frac{1}{2m}\left|\nabla \tilde a-i\left(\frac{1}{2}{\vec A}^{\rm ext}+\vec{\mathcal{A}}\right)\tilde a+i\vec\alpha\tilde a\right|^2,\\
\mathcal{L}_b=&i\tilde b^*\left[\partial_0-i\left(\frac{1}{2}A^{\rm ext}_0-\mathcal{A}_0\right)+i\beta_0\right]\tilde b\nonumber\\
&-\frac{1}{2m}\left|\nabla \tilde b-i\left(\frac{1}{2}{\vec A}^{\rm ext}-\vec{\mathcal{A}}\right)\tilde b+i\vec\beta \tilde b\right|^2,
\end{align}
and,
\begin{align}
\mathcal{L}_{CS}=&\frac{1}{4\pi}\epsilon^{\mu\nu\lambda}\left[\alpha_\mu\partial_\nu\beta_\lambda+\beta_\mu\partial_\nu\alpha_\lambda\right],\\
\mathcal{L}_{\mathcal{A}}=&-\frac{1}{4e^2}(\partial\mu\mathcal{A}_\nu-\partial_\nu\mathcal{A}_\mu)^2.
\end{align}
Here $\alpha,\beta$ are the statistical Chern-Simons (CS) fields that implement the mutual flux attachment. 
$\mathcal{L}_{\rm int}$ represent quadratic and quartic terms for the composite bosons that can be tuned to condense them. 
$\mathcal{A}$ represents the internal dynamic gauge field,
and $A^{\rm ext}$ represents an external test field that couples to the $S^z$ charge of the microscopic models.

When $\tilde a,\tilde b$ condense to stabilize the $U(1)$ SPT, the CS statistical gauge fields are locked as
\begin{align}
\alpha=\frac{1}{2}A^{\rm ext}+\mathcal{A};\quad \beta=\frac{1}{2}A^{\rm ext}-\mathcal{A}.
\end{align}
Thus the CS term becomes the action of CSL,
\begin{align}
\mathcal{L}_{CS}=\frac{1}{4\pi}\epsilon^{\mu\nu\lambda}\left[\frac{1}{2}A^{\rm ext}_\mu\partial_\nu A^{\rm ext}_\lambda-2\mathcal{A}_\mu\partial_\nu\mathcal{A}_\lambda\right].
\end{align}
The last term on the right, namely the CS term for the emergent gauge field, prohibits the creation of instantons (monopoles) in $2+1$ dimensions, hence the system is deconfined \cite{Fradkinbook};
and along with the Maxwell term for $\mathcal{A}$ provides a mass $m\sim e^2$ for the photon and hence gaps it out and the low energy theory is given by the first term which gives a Hall conductivity (of $S^z$) $\sigma_{xy}=1/2$. 

It is worth elaborating on two important points about the relation between the $U(1)$ SPT and CSL. 
Firstly,  the $U(1)$ SPT requires one global $U(1)$ symmetry, either the $U(1)$ charge or the $U(1)$ pseudospin, to protect it.
The $U(1)$ pseudospin symmetry ($n_a-n_b$) corresponds to the local gauge structure in the context of the CSL phase, which can never be broken hence protects the $U(1)$ SPT.
On the contrary, breaking the $U(1)$ charge corresponds to breaking global $S^z$ conservation of the CSL. As the CSL is  topologically ordered, it is robust against such symmetry breaking.
Secondly, we need to understand how to match edge modes, namely the two counter-propagating edge modes of the $U(1)$ SPT and the single chiral edge mode of the CSL.
The two edge modes of the $U(1)$ SPT are a left-mover carrying charge and a right-mover carrying pseudospin.
Note that the pseudospin mode is coupled to the dynamical gauge field $\mathcal A$ and will thus be removed after integrating out $\mathcal A$ as it acquires a Higgs mass \cite{Wen1991}. 
As a result, only the charge mode is left which becomes the chiral  mode of the CSL.
This edge mode does not require the global $S^z$ conservation to protect it.
%
%

\paragraph{Numerical verification.} Using numerical techniques, we can show that the original kagome system Eq.~(\ref{eq:Ham_B}) has indeed a CSL ground state---in keeping with our  gauge theory analysis. 
Similar  studies have been discussed in detail for other CSL elsewhere \cite{He2015,He2014,ssgong14,Bauer2014} and thus we will here only briefly review our numerical results.     
In particular, we use the iDMRG method \cite{McCulloch2008} to obtain the two-fold topological degenerate ground states \cite{He2014a}, calculate fractional Hall conductance \cite{Hall_conductance}, obtain the braiding statistics (via modular matrices \cite{Zhang2012,Cincio2013,Wen1989}), and probe chiral gapless edge mode from entanglement spectra \cite{Li2008} (see supplemental materials).
All those calculations validate the characteristic topological properties of a CSL which are robust against finite size effects, do not involve any data extrapolation, and hence are quite reliable.

\begin{figure}
\includegraphics[width=0.45\textwidth]{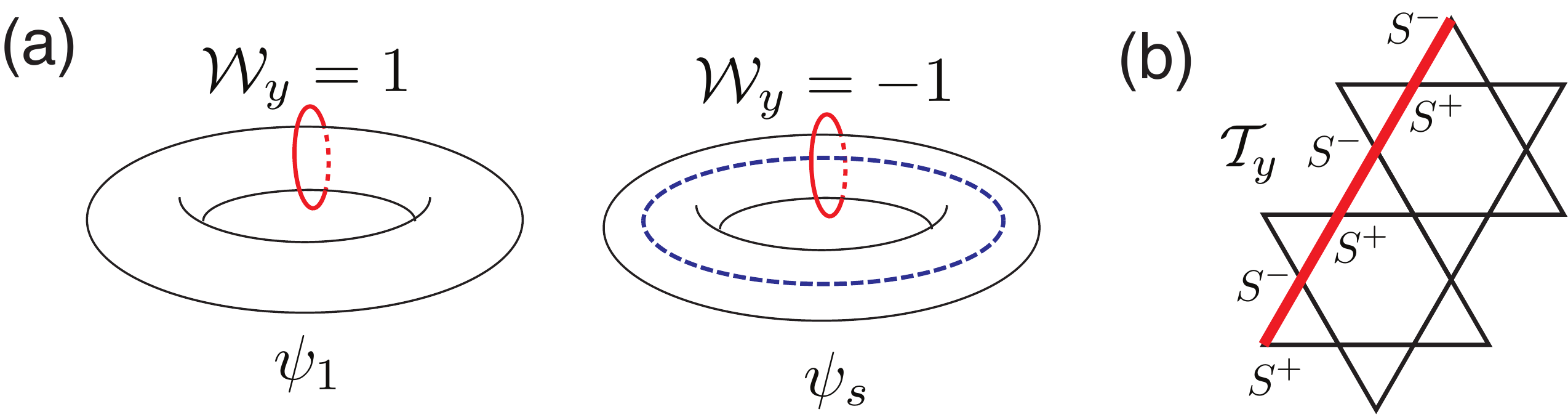} \caption{\label{fig:kagome_CSL} (a) Topological degenerate ground states distinguished by the Wilson loop operator. (b) Approximation of the Wilson loop operator of the microscopic kagome model.}
\end{figure}

Interestingly, based on the lattice gauge theory, we can  identify the Wilson loop operator $\mathcal W_y$ in the microscopic spin model. 
The Wilson loop operator is a global operator that distinguishes different topological degenerate ground states, and can be obtained as follows: first create a pair of spinons, then wind them around the torus or infinite cylinder, and finally annihilate them. 
Then the expectation value of the Wilson loop operator for the two topologically degenerate ground states ($\psi_{1(s)}$) is $\mathcal W_y\sim \pm 1$ [Fig. \ref{fig:kagome_CSL}(a)]. 
From our microscopic lattice gauge theory, the Wilson loop corresponds to a gauge flux around the non-contractible loop along the $y$ direction (Fig. \ref{fig:kagome_CSL}(b)):
\begin{equation}
\mathcal W_y=\exp(i\sum_{\mathcal {NC}_y} \mathcal A_{ij})= \mathcal P \left( \prod_{\mathcal {NC}_y} S^+_i S^-_j \right) \mathcal P = \mathcal P\,\mathcal T_y \, \mathcal P
\end{equation}
$\mathcal P$ represents a projection into the classically degenerate manifold. 
Technically, the projection can be treated as a renormalization, and approximated by: $\langle \psi | \mathcal W_y |\psi \rangle  \approx \langle \psi | \mathcal T_y  |\psi \rangle/\langle \psi | \mathcal T_y^\dag \mathcal T_y |\psi \rangle$.
Numerically we find that $ \langle \psi | \mathcal T_y  |\psi \rangle/\langle \psi | \mathcal T_y^\dag \mathcal T_y |\psi \rangle\sim \pm 0.8$, which is close to the above theoretical expectation.

\paragraph{Anisotropic kagome XXZ model:} We finally discuss the relevance of our findings for the extended kagome XXZ model \cite{He2015}
\begin{align}
&H_{XXZ} =J_z \sum_{\langle p q\rangle} S_p^z S_q^z+\frac{J_{xy}}{2}\sum_{\langle p q\rangle}  (S^+_p S^-_q +h.c.) \nonumber  
\\ &+ \frac{J'_{xy}}{2}\sum_{\langle\langle p q\rangle\rangle} (S^+_p S^-_q +h.c.)  + \frac{J'_{xy}}{2}\sum_{\langle\langle\langle p q\rangle\rangle\rangle} (S^+_p S^-_q +h.c.),
\label{eq:XXZ}
\end{align}
with first neighbor $\langle  \rangle$ $XXZ$ interactions,  second- $\langle\langle \rangle\rangle$ and third- $\langle\langle\langle \rangle\rangle\rangle$ neighbor $XY$ interactions (see Fig. \ref{fig:kagome_honeycomb}).

\begin{figure}
\includegraphics[width=0.45\textwidth]{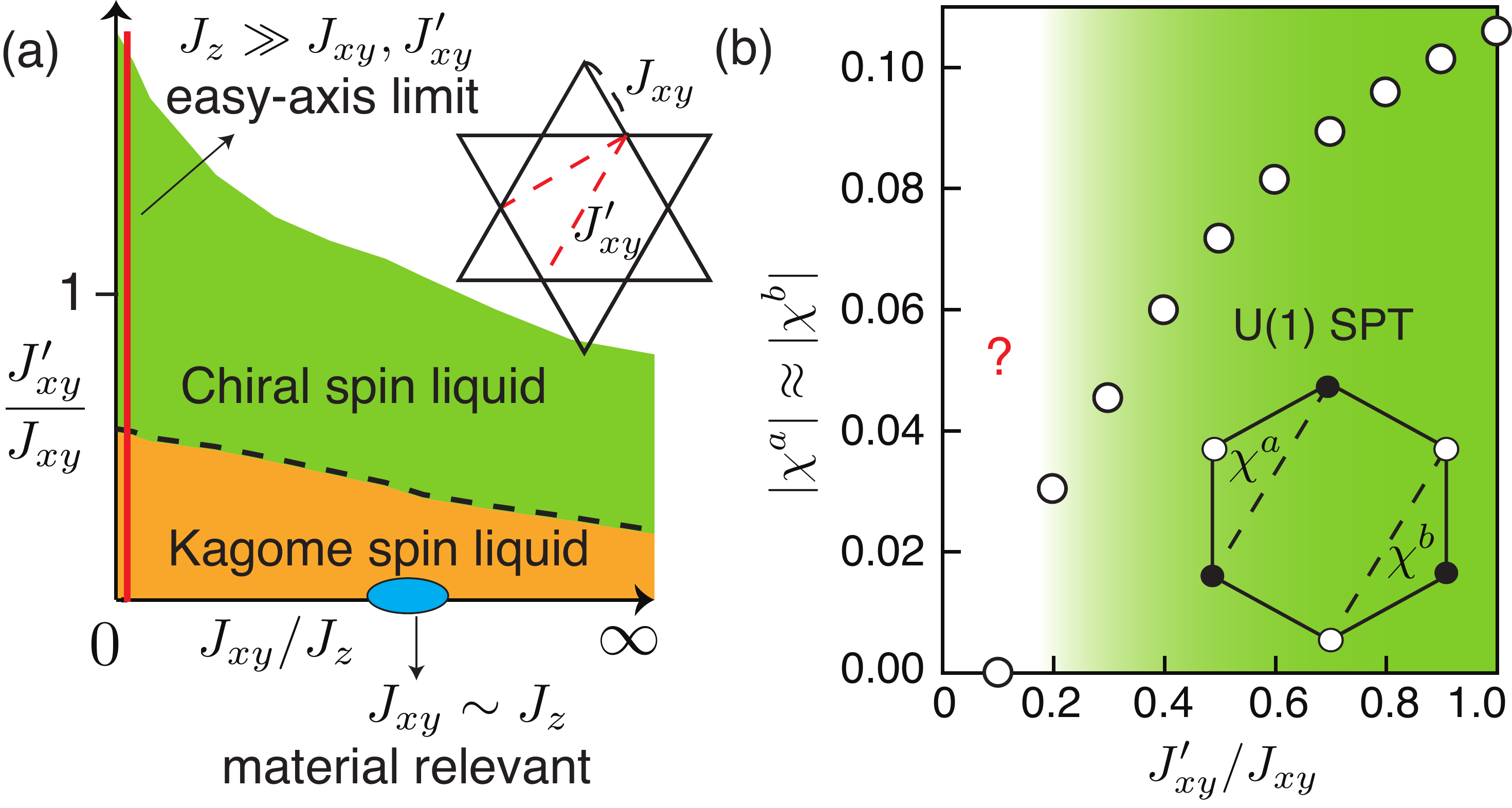} 
\caption{(a) Phase diagram of the $XXZ$ kagome model in Eq.~(\ref{eq:XXZ}).
(b) Order parameter for spontaneous time-reversal symmetry breaking in the gMF Hamiltonian, here width of cylinder is $W_y=8$ sites.  \label{fig:XXZ}}
\end{figure}

The general phase diagram  Fig. \ref{fig:XXZ}(a) \cite{He2015} shows an extended chiral spin liquid phase that spontaneously breaks time reversal symmetry, for sufficiently strong second and third neighbor  $XY$ interactions $J_{xy}\sim J_{xy}'$.
Analogous to the derivation for the chiral model Eq. (\ref{eq:Ham_B}) above, we find the effective lattice gauge Hamiltonian for the limit $J_{xy}, J_{xy}'\ll J_{z}$ which reads (with unimportant factors omitted)
\begin{align} \label{eq:LGT_HamA}
&H_{\textrm{XXZ}}^{\textrm{LGT}}=J_{xy}  \Big[\sum_{\langle \langle ij \rangle\rangle}e^{i \mathcal  A_{ij}}  a^\dag_i a_j + \sum_{\langle \langle kl \rangle\rangle} e^{i\mathcal A_{lk}} b^\dag_k b_l + h.c. \Big] \nonumber
 \\ & + J'_{xy} \sum_{\langle i k\rangle, \langle j l \rangle \in \hexagon}  \left[(e^{i\mathcal A_{ik} } a^\dag_i b_k^\dag)( e^{ i\mathcal A_{lj}} b_l a_j) +h.c.\right].
\end{align}

Again we use the gMF approach to solve this Hamiltonian. 
An interesting observation is that its gMF Hamiltonian can be written as the product of correlated hopping terms (see supplemental materials):
\begin{align}
\widetilde H_{XXZ}=&-J_{xy}\bigg[\sum_{ ijm, k} \chi_{im,k}^a \chi_{mj,k}^a +\sum_{kln,j} \chi_{kn,j}^b \chi_{nl,j}^b+h.c.\bigg] \nonumber \\ 
& -J'_{xy} \sum_{\langle i k\rangle, \langle j l \rangle \in \hexagon}  \chi^a_{ij, k} \chi^b_{kl,j}.
\end{align}
We use a generalized form of correlated hopping,
\begin{align}
\chi_{ij,l}^a&=i (2n_l^b-1) e^{i\mathcal A_{ij}^0} a_i^\dag a_j,~~~ \chi_{ji,l}^a=(\chi_{ji,l}^a)^\dag,  \label{eq:g_cpa} 
\end{align}
and similarly for $\chi^b$. 
Here we do not require  site $l$ to be the nearest neighbor of site $i$ and $j$.
Thus it is reasonable to expect that a finite $J_{xy}'$ could induce a spontaneous time-reversal symmetry breaking with $\chi^a, \chi^b\neq 0$, and result in a $U(1)$ SPT phase (in the gMF Hamiltonian)  as in the correlated hopping model \cite{He2015b}.
With the help of iDMRG simulations we find that this is indeed true for systems with sufficiently large $J'_{xy}$  (see supplemental materials). For example the order parameter $\chi^a, \chi^b$ for the time-reversal symmetry breaking is shown in Fig. \ref{fig:XXZ} (b).
Therefore, the CSL in the kagome XXZ model can also be explained as  a gauged $U(1)$ SPT  following our construction. 

\paragraph{Conclusion:} We have achieved a theoretical understanding of recent numerically discovered chiral spin liquids in kagome antiferromagnets, which turns out to be a gauged $U(1)$ symmetry protected topological phase. 
It  is a rare example that  starting from a microscopic model, a (non-$ Z_2$)  spin liquid  phase  can be described in a controlled way
in two spatial dimensions, via a faithful gauge theoretic model accounting for the QSL as its  deconfined phase. 
This framework might be applicable to numerous interesting problems, such as solving the precise nature of the kagome spin liquid as well as realizing exotic topological phases through the ``gauging" procedure in related kagome systems or doped quantum dimer models. %

\emph{Acknowledgement.}---We thank M. Oshikawa and T. Senthil for discussions. This work was supported by the Deutsche Forschungsgemeinschaft (DFG) through the collaborative research centre SFB 1143.

\newpage

\begin{widetext}
\noindent{\bf \large Supplementary Materials of ``Kagome chiral spin liquid as a gauged $U(1)$ symmetry protected topological phase"}
\end{widetext}

\section{Derivation of the lattice gauge Hamiltonian}
Now we use the lattice gauge mapping we outlined above to derive the lattice gauge Hamiltonian of the two spin models.
The first model is
\begin{align}  \label{eq:HamB_sub}
H_{\text{chiral}}= J_z \sum_{\langle p q\rangle} S_p^z S_q^z+\lambda\sum_{p,q,r\in \bigtriangledown, \bigtriangleup} \vec S_r \cdot (\vec S_p \times \vec S_q),
\end{align}
where $J_z\gg \lambda$, the three-spin term is called scalar chirality.

For a down-triangle of the kagome labelled by the sites $p, q$ and $r$, the scalar chirality is given by (we have traversed the triangle in an anti-clock wise direction)
\begin{align}
{\vec S}_p\cdot({\vec S}_q\times {\vec S}_r)
&=\frac{i}{2}\left[S_r^z(S^+_pS^-_q-S_p^-S^+_q)+ {\rm cyclic~perm.}\right]
\end{align} 
For example the term
\begin{align}
\frac{i}{2}S_r^z(S^+_pS^-_q-S_p^-S^+_q)
\end{align}
can only yield non-zero matrix element within the easy-axis ground state sector when $S^z_p$ and $S^z_q$ are anti-parallel. In that case (assume an up-triangle)
\begin{align}
n_k^b=S^z_p+S^z_q+S^z_r+\frac 1 2=S^z_r+\frac 1 2.
\end{align}
On the other hand, $S^+_p S_q^-$ will  create spinon at site $i$, and annihilate spinon at site $j$.
Therefore the above term becomes, after using the hard-core boson representation,
\begin{align}
\frac{1}{4}(2n^b_k-1)\left[e^{i(\mathcal{A}_{ij}+\pi/2)}\hat a^\dagger_i \hat a_j+{\rm h.c.}\right]
\end{align}
This is nothing but a correlated hopping term in presence of the gauge field which is dynamic in this case.

Now let us consider the XXZ model, 
\begin{align}\label{eq:XXZ_sub}
&H_{XXZ} =J_z \sum_{\langle p q\rangle} S_p^z S_q^z+\frac{J_{xy}}{2}\sum_{\langle p q\rangle}  (S^+_p S^-_q +h.c.) \nonumber  
\\ &+ \frac{J'_{xy}}{2}\sum_{\langle\langle p q\rangle\rangle} (S^+_p S^-_q +h.c.)  + \frac{J'_{xy}}{2}\sum_{\langle\langle\langle p q\rangle\rangle\rangle} (S^+_p S^-_q +h.c.).
\end{align}
It is also straightforward to derive the lattice gauge Hamiltonian. First neighbor $XX$ interaction yields
\begin{equation}
\sum_{\langle \langle ij \rangle\rangle}e^{i \mathcal  A_{ij}}  a^\dag_i a_j + \sum_{\langle \langle kl \rangle\rangle} e^{i\mathcal A_{lk}} b^\dag_k b_l + h.c. 
\end{equation}
The second and third neighbor interactions yield:
\begin{equation}
\sum_{\langle i k\rangle, \langle j l \rangle \in \hexagon}  \left[(e^{i\mathcal A_{ik} } a^\dag_i b_k^\dag)( e^{ i\mathcal A_{lj}} b_l a_j) +h.c.\right].
\end{equation}

\section{$U(1)$ SPT phase}

\begin{figure}
\includegraphics[width=0.3\textwidth]{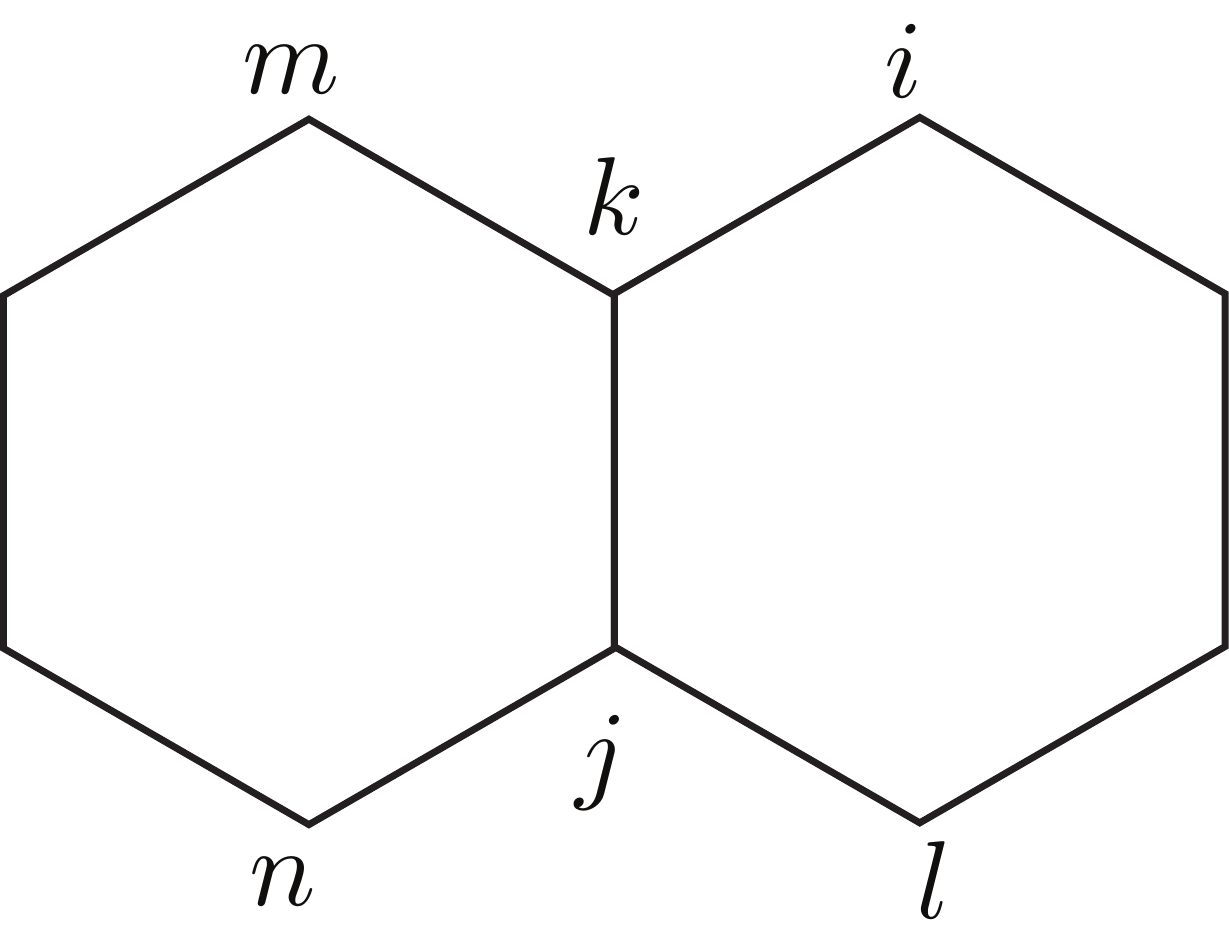} \caption{\label{fig:notation}Figure for the form of interactions.}
\end{figure}

\subsection{Gauge mean field Hamiltonian of XXZ model}
The gauge mean-field Hamiltonian for the XXZ model is
\begin{align}
\widetilde{H}_{XXZ}&=J_{xy}  \Big[\sum_{\langle \langle ij \rangle\rangle}e^{i \mathcal  A_{ij}^0}  a^\dag_i a_j + \sum_{\langle \langle kl \rangle\rangle} e^{i\mathcal A_{lk}^0} b^\dag_k b_l + h.c. \Big] \nonumber
 \\ & + J'_{xy} \sum_{\langle i k\rangle, \langle j l \rangle \in \hexagon}  \left[(e^{i\mathcal A_{ik}^0 } a^\dag_i b_k^\dag)( e^{ i\mathcal A_{lj}^0} b_l a_j) +h.c.\right].
 \label{eq:spinon_HamA}
\end{align}
The background flux here is $\phi_{\mathcal A^0}=\sum_{\hexagon} \mathcal A=\pi$ for each hexagon.

It is helpful to define a generalized form of correlated hopping as
\begin{align}
\chi_{ij,l}^a&=i (2n_l^b-1) e^{i\mathcal A_{ij}^0} a_i^\dag a_j,~~~ \chi_{ji,l}^a=(\chi_{ji,l}^a)^\dag,  \label{eq:g_cpa} \\
\chi_{kl,m}^b&=i (2n_m^a-1) e^{i\mathcal A_{kl}^0} b_k^\dag b_l,~~~ \chi_{lk,m}^b=(\chi_{kl,m}^b)^\dag, \label{eq:g_cpb}
\end{align}
where we donot require that  site $l$ to be the nearest neighbor of site $i$ and $j$.
Under the time-reversal symmetry, we have
\begin{equation}
\mathcal T: ~~~~ ~~~~ \chi \rightarrow -\chi.
\end{equation}
We expect that this generalized correlated hopping will favor the mutual flux attachment and stabilizes a $U(1)$ SPT phase.

Let us rewrite $\widetilde H_{XXZ}$ of Eq.~(\ref{eq:spinon_HamA}) in terms of the generalized correlated hoppings, Eq.~(\ref{eq:g_cpa}) and Eq.~(\ref{eq:g_cpb}). Fig. \ref{fig:notation} shows the notation we use in the following. For the two body  hopping term, we have
\begin{align}
e^{i\mathcal A_{ij}^0} a^\dag_i a_j&=e^{i\mathcal A_{im}^0+i\mathcal A_{mj}^0} a^\dag_i a_j (a^\dag_m a_m+a_m a^\dag_m) (2n_k^b-1)^2 \nonumber \\
&=-\chi_{im,k}^a \chi_{mj,k}^a- \chi_{mj,k}^a\chi_{im,k}^a.
\end{align}
Here we have used the property of hard-core boson, such that $a^\dag_m a_m+a_m a^\dag_m=1$, $(2n_k^b-1)^2=1$.

For the four body interactions $(e^{i\mathcal A_{ik}^0 } a^\dag_i b_k^\dag)( e^{ i\mathcal A_{lj}^0} b_l a_j) $, we have
\begin{align}
&(e^{i\mathcal A_{ik}^0} a^\dag_i b^\dag_k ) (e^{i\mathcal A_{lj}^0} a_j b_l) =(e^{i\mathcal A_{ij}^0} a^\dag_i a_j) (e^{i\mathcal A_{lk}^0} b^\dag_k b_l) \nonumber \\
&= (e^{i\mathcal A_{ij}^0} a^\dag_i a_j)  (2n_k^b-1) (2n_j^a-1) (e^{i\mathcal A_{lk}^0} b^\dag_k b_l) \nonumber \\
&=-\chi^a_{ij, k} \chi^b_{kl,j},
\end{align}
where we explore the  property of hard-core boson again, $(2n_k^b-1) b^\dag_k=b_k^\dag$ and $a_j (2n_j^a-1)=a_j$.

Therefore,
\begin{align}
\widetilde H_{XXZ}=&-J_{xy}\bigg[\sum_{ ijm, k} \chi_{im,k}^a \chi_{mj,k}^a +\sum_{kln,j} \chi_{kn,j}^b \chi_{nl,j}^b+h.c.\bigg] \nonumber \\ 
& -J'_{xy} \sum_{\langle i k\rangle, \langle j l \rangle \in \hexagon}  \chi^a_{ij, k} \chi^b_{kl,j}.
\end{align}
Now it is clear that why it is reasonable to expect that the $\widetilde H_{XXZ}$ in Eq.~(\ref{eq:spinon_HamA}) host a $U(1)$ SPT phase, which breaks time-reversal symmetry spontaneously with the emergence of a finite order parameter $\chi^a, \chi^b\ne 0$.

\subsection{Numerical verification of $U(1)$ SPT phase}
\begin{figure}
\includegraphics[width=0.45\textwidth]{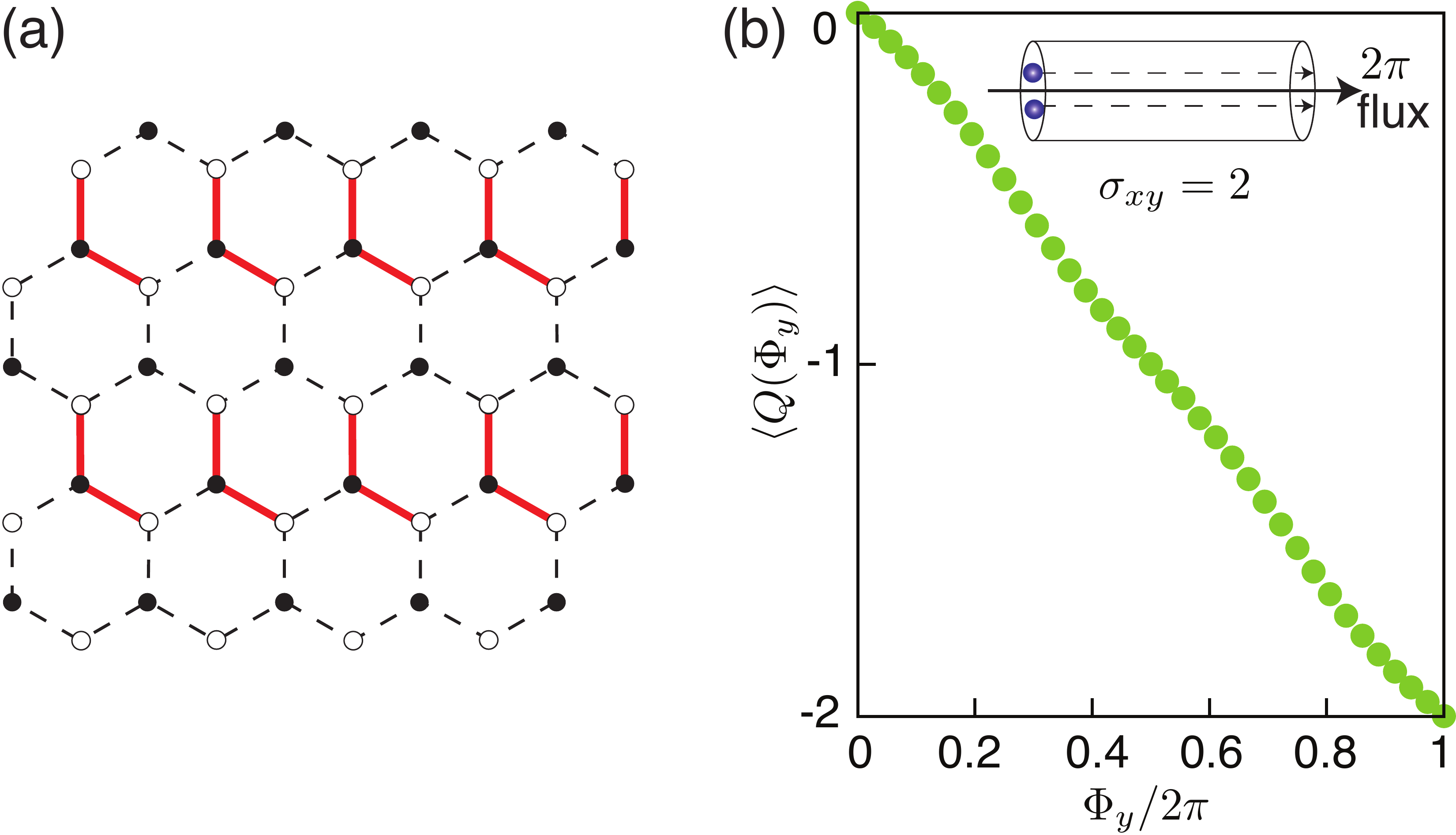} \caption{(a) Gauge chosen for numerical simulation. The gauge field on the red solid link is $\pi$, on the dashed link is $0$. (b) Quantized charge pumping for the $U(1)$ SPT phase, here $J'=1$, the width is $W_y=8$ sites. \label{fig:SPT_sup}}
\end{figure}

In the last section, we have given an intuitive argument that why $\widetilde H_{XXZ}$ in Eq.~(\ref{eq:spinon_HamA}) has a $U(1)$ SPT ground state with spontaneous time-reversal symmetry breaking. 
Here we will use DMRG to prove that this is indeed true.
To implement the simulation, we wrap the system around a cylinder,  double the unit cell to $4$ sites (due to the background  flux  $\pi$ in each hexagon), and choose a gauge as shown in Fig. \ref{fig:SPT_sup}(a).
We have calculated systems of width $W_y=8, 12, 16$ sites (corresponding to $L_y=2,3,4$ unit cells), which all give the same $U(1)$ SPT phase.

Similar as what we did in Ref. \cite{He2015b}, we will discuss in detail two characteristic fingerprints of the $U(1)$ SPT phase to establish its existence: (i) the ground state has a quantized Hall conductance $|\sigma_{xy}|=2$ ;  (ii) the ground state has two counter-propagating gapless edge modes.

The $U(1)$ SPT phase is also called the bosonic integer quantum Hall state, hence the quantized Hall conductance is its hallmark. In contrast to fermionic systems, the Hall conductance $\sigma_{xy}$ of a $U(1)$ SPT state is always quantized to an even number \cite{Senthil2013}.  Numerically, we can use an adiabatic flux insertion to measure the Hall conductance $\sigma_{xy}$: $2\pi$ flux insertion on a cylinder will pump $\sigma_{xy}$ particles from the left edge to the right edge of cylinder.  Flux insertion can be implemented in the Hamiltonian by twisting the boundary condition in the infinite DMRG algorithm \cite{ He2014a}: the bosons hopping around the cylinder pick up a flux $\Phi_y$.  
The Hall conductance can then be written as \cite{Hall_conductance}:
\begin{align}
\sigma_{xy}&=\int_{0}^{2\pi}  [\partial_{\Phi_y} \langle Q(\Phi_y)\rangle] d \Phi_y, \\
 \langle Q(\Phi_y)\rangle &= \sum_{i} \lambda_i(\Phi_y) Q_i(\Phi_y),
\end{align}
where $\lambda_i(\Phi_y)$ are the eigenvalues of the reduced density matrix when flux $\Phi_y$ is inserted, $Q_i(\Phi_y)$ is the corresponding $U(1)$ quantum number.

The existence of the symmetry protected gapless edge modes is another hallmark of the SPT phase.  
The $U(1)$ SPT state has two counter-propagating edge modes, which can be identified as a charge mode that carries charge with no pseudospin, and  a pseudospin mode that carries pseudospin with no charge. 
Thus, as long as one $U(1)$ symmetry (charge conservation or pseudospin conservation) is preserved, backscattering between the two edge modes is prohibited \cite{Senthil2013}. 

Similar as fractional quantum Hall states \cite{Wen_book}, one can use an  Abelian Chern-Simons theory with the $K$-matrix $K=\left( \begin{matrix} 0 & 1 \\ 1 &0 \end{matrix} \right)$ \cite{Lu2012, Senthil2013} to describe the $U(1)$ SPT phase and its edge modes:
\begin{equation}
\mathcal L=-\frac{1}{4\pi} (K_{\alpha\beta} \partial_t \phi_\alpha \partial_x \phi_\beta+V_{\alpha\beta} \partial_x \phi_\alpha \partial_x \phi_\beta),
\end{equation}
where $\alpha, \beta=A, B$ and $1/2\pi \partial_x \phi_\alpha$ gives the density of the corresponding species of bosons, and $V_{\alpha\beta}$ is the velocity matrix. 
To diagonalize the above Lagrangian, we introduce the charge and pseudospin modes $\phi_{c(s)}=(\phi_a\pm \phi_b)/\sqrt 2$. 
We can now obtain the edge Hamiltonian and the corresponding momentum operator:
\begin{equation}
H=\frac{2\pi}{L_y}(v_c L_0^c+v_s L_0^s), \quad\quad P=\frac{2\pi}{L_y}( L_0^c- L_0^s), \label{eq:two_modes}
\end{equation}
with
\begin{equation}
L_0^{c(s)}=\frac{(\Delta N_a\pm \Delta N_b)^2}{4}+\sum_{m=1}^{\infty} m n_m^{c(s)}. \label{eq:counting}
\end{equation}
Here, $L_y$ is the length of the 1D edge; $\Delta N_{a(b)}$ is the change in the particle number of $a(b)$ boson relative to the ground state; $\{ n_m^{c(s)}\}$ is the set of non-negative integers describing oscillator modes. As compared to  FQH states with only one chiral mode, Eq.~(\ref{eq:two_modes}) shows two counter propagating modes.

\begin{figure}
\includegraphics[width=0.48\textwidth]{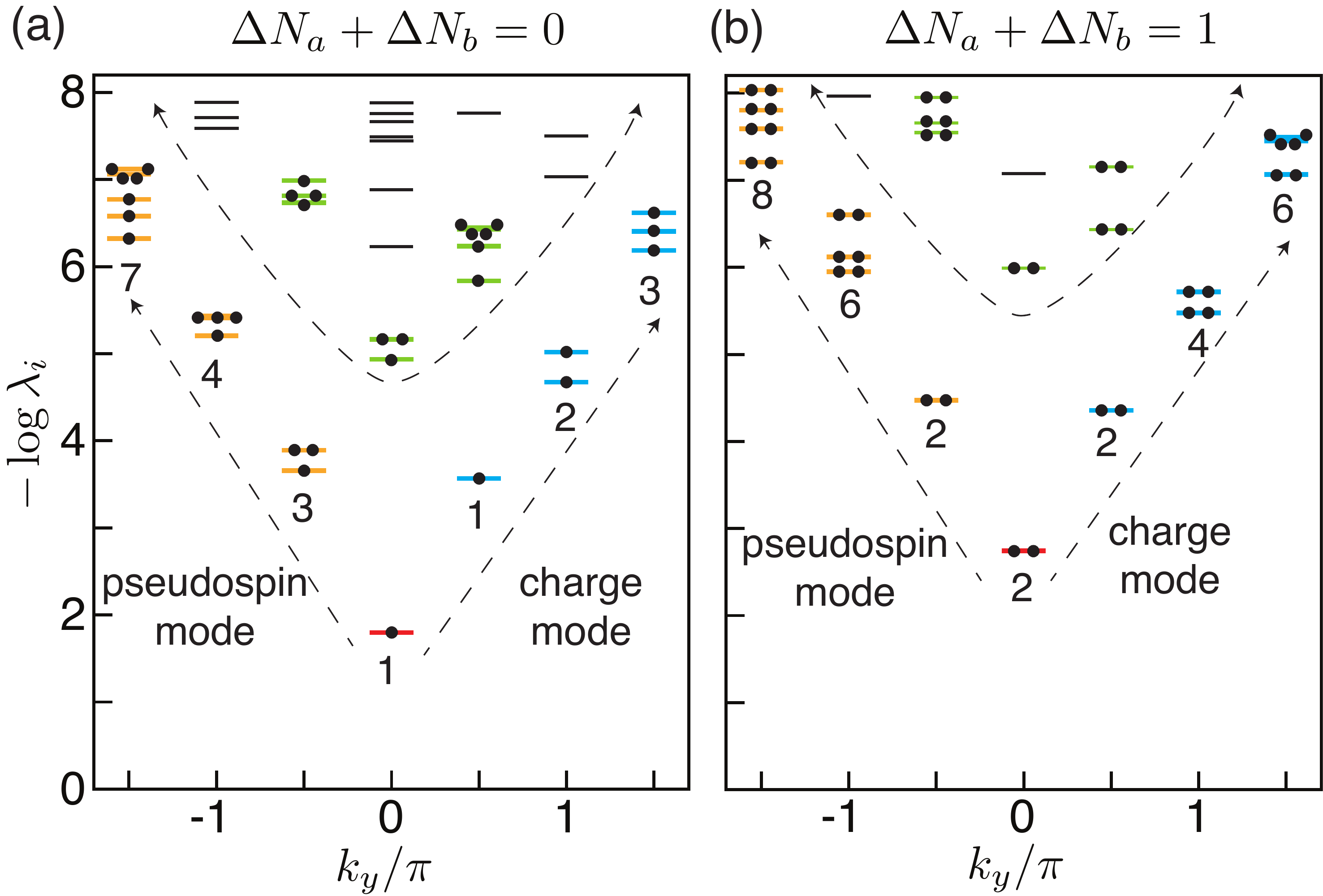}\caption{\label{fig:SPT_ES} The entanglement spectra versus momentum $k_y$: (a) charge sector $\Delta N_a+\Delta N_b=0$. (b) charge sector $\Delta N_a+\Delta N_b=1$. The simulation is carried on an  infinite cylinder of width $L_y=4$ unit cells ($W_y=16$ sites), $J'=1$. }
\end{figure}

Numerically, we can use the entanglement spectra as a probe of the edge modes \cite{Li2008}. 
The numerical results from the DMRG simulation are shown in Fig. \ref{fig:SPT_ES}. 
We have plotted two different cases that correspond to the $U(1)$ charge sector $\Delta N_a+\Delta N_b=0$ and $\Delta N_a+\Delta N_b=1$. 
The two counter propagating edge modes are clearly seen, and their counting in each sector from our numerics agrees well with the theoretical expectation \cite{He2015b}. 
%
\section{Chiral spin liquid}

Here we provide the numerical results for the chiral spin liquid in the three-spin model,
\begin{align}
H_{\textrm{chiral}}= J_z \sum_{\langle p q\rangle} S_p^z S_q^z+\lambda\sum_{p,q,r\in \bigtriangledown, \bigtriangleup} \vec S_r \cdot (\vec S_p \times \vec S_q),  \label{eq:Ham_B_sup}
\end{align}
with $J_z\gg \lambda>0$.
 Similar as the $U(1)$ SPT phase, the chiral spin liquid phase has a large gap and short correlation length, the iDMRG is a very reliable method to study them. 

Again, we wrap the kagome lattice on a cylinder and use the infinite DMRG method to solve its ground state. To verify the ground state is indeed a CSL phase, we have numerically proved that the it has all the topological properties of a CSL phase, including the two-fold topological degeneracies, the quantized (fractional) charge pumping, the fractional statistics and a gapless chiral edge modes.

\begin{figure}
\includegraphics[width=0.49\textwidth]{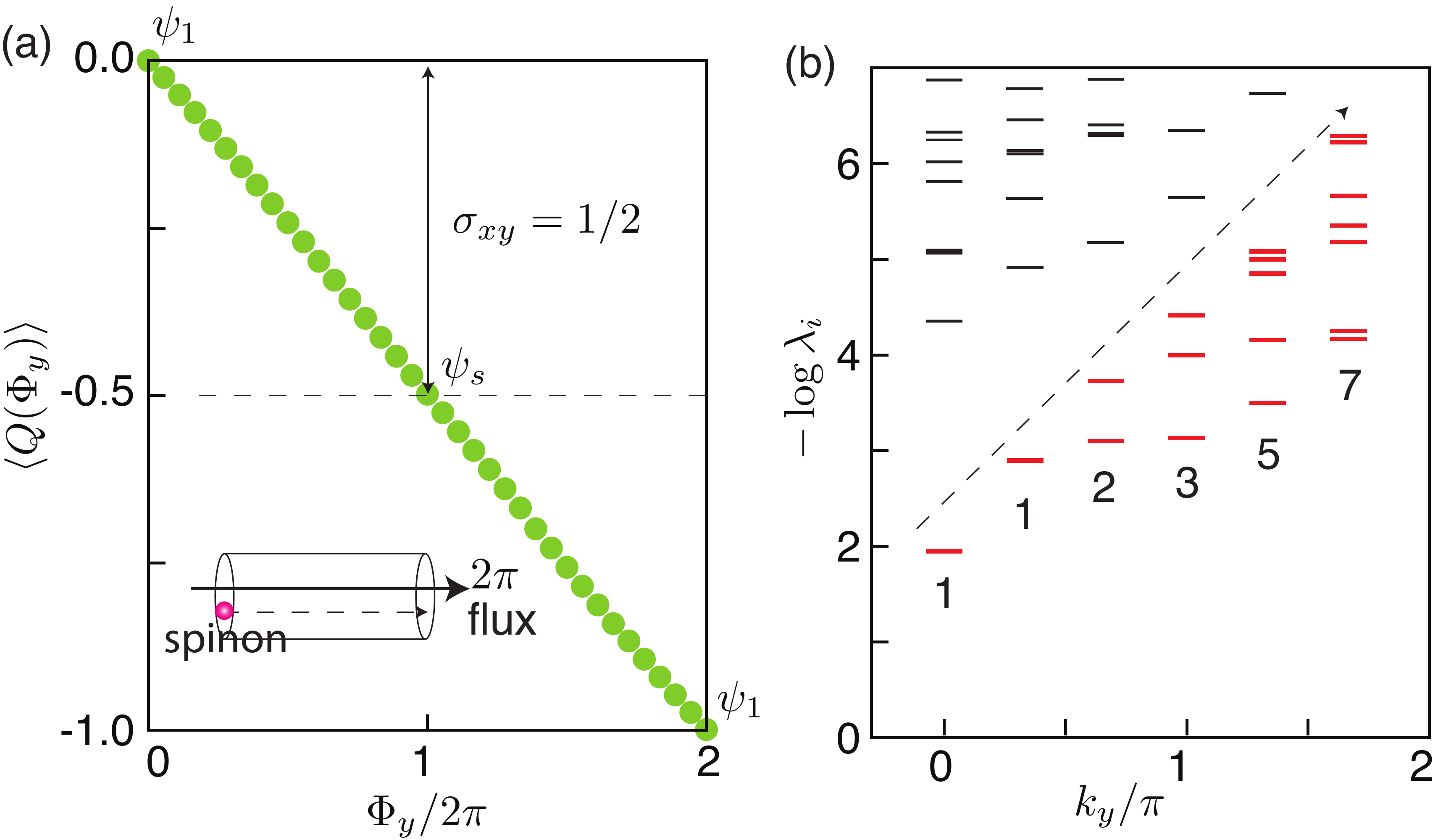} \caption{(a) Quantized charge pumping. After inserting $2\pi$ flux, a spinon (half spin quantum number) is pumped, and the topological degenerate ground state $\psi_1$ evolves into the other degenerate state $\psi_s$. (b) Entanglement spectra in the spin sector $\Delta S^z=0$, the cylinder's width is $L_y=6$ unit cells. \label{fig:CSL}}
\end{figure}

Firstly, we perform a $2\pi$ flux insertion in our numerical experiment (shown in Fig. \ref{fig:CSL}(a)), and find that a spinon (carries $1/2$ spin quantum number) is pumped from the left edge to the right edge of the cylinder, meanwhile the topologically degenerate ground state ($\psi_1$) adiabatically evolves into the other topologically degenerate ground state $\psi_s$. With these two topological degenerate ground states, we can calculate the modular matrix \cite{Wen1990,Zhang2012, Cincio2013}, which gives:
\begin{align}
\mathcal S&=\frac{1}{\sqrt 2}\left( \begin{matrix} 1 & 1 \\ 1 & -1 \end{matrix} \right)+ o(10^{-2}),
\end{align}
and
\begin{align}
\mathcal U&=e^{-i (2\pi/24)} \left( \begin{matrix} 1 & 0 \\ 0 & i \end{matrix} \right)\times o(10^{-2}).
\end{align}
The modular matrix fully characterizes the topological order of a CSL phase, from which we can extract the fractional statistics, the fusion rule and the quantum dimension. For example, from the $\mathcal S$ matrix, we know the fractional statistics obeyed by the spinon: one spinon encircling another spinon will give rise to a non-trivial phase factor $-1$.

Furthermore, we use the entanglement spectra to probe the gapless edge mode of the CSL phase. As shown in Fig. \ref{fig:CSL} (b), the entanglement spectra show one chiral edge mode with positive momentum, and it agrees with the counting rule $1, 1, 2, 3, 5, \cdots$.   

\end{document}